\documentclass[aps,twocolumn,showpacs,superscriptaddress,floatfix]{revtex4-1}
\usepackage{graphicx,amsfonts,amssymb,amsmath,mathrsfs,hyperref,bm}

\usepackage{epstopdf}

\newif\ifhyper
\hypertrue
\ifhyper
\hypersetup{
   citecolor = {red},
   colorlinks = {true}, 
   linkcolor = {blue},
   urlcolor = {blue} 
}
\fi

\begin{document}

\title{
Quantum phase diagram of two-dimensional transverse field Ising model: \\
unconstrained tree tensor network and mapping analysis}

\author{M. Sadrzadeh}
\affiliation{Department of Physics, Sharif University of Technology, P.O.Box 11155-9161, Tehran, Iran}

\author{R. Haghshenas}
\affiliation{Department of Physics and Astronomy, California State University, Northridge, California 91330, USA}

\author{A. Langari}
\affiliation{Department of Physics, Sharif University of Technology, P.O.Box 
11155-9161, Tehran, Iran}
\email{langari@sharif.edu}

\begin{abstract}
We investigate the ground-state phase diagram of the frustrated transverse field Ising (TFI) model on the checkerboard lattice (CL), which consists of N\'{e}el, collinear, quantum paramagnet and plaquette-valence bond solid (VBS) phases. We implement a numerical simulation that is based on the recently developed unconstrained tree tensor network (TTN) ansatz, which
systematically improves the accuracy over the conventional methods as it exploits the internal gauge selections. At the highly frustrated region ($J_2=J_1$), we observe a second order phase transition from plaquette-VBS state to paramagnet phase at the critical magnetic field, $\Gamma_{c}=0.28$, 
with the associated critical exponents $\nu=1$ and $\gamma\simeq0.4$, which are obtained within
the finite size scaling analysis on different lattice sizes $N=4\times 4, 6\times 6, 8\times8$.  
The stability of plaquette-VBS phase at low magnetic fields is examined by spin-spin correlation
function, which verifies the presence of plaquette-VBS at $J_2=J_1$ and rules out the existence of 
a N\'{e}el phase. In addition, our numerical results suggest that the transition from 
N\'{e}el (for $J_2<J_1$) to plaquette-VBS phase is a deconfined phase transition. 
Moreover, we introduce a mapping, which renders the low-energy effective theory of TFI on CL 
to be the same model on $J_1-J_2$ square lattice (SL).
We show that the plaquette-VBS phase of the highly frustrated point $J_2=J_1$ on CL
is mapped to the emergent string-VBS phase on SL at $J_2=0.5J_1$. 
\end{abstract}


\maketitle
\section{Introduction}
\label{introduction}
Quantum phases of matter without magnetic long-range order have become an interesting 
field of research in recent years. Frustrated magnetic systems are one of the best candidates to bring about such phases like spin-ice materials or spin liquids \cite{Harris:1997,Bramwell:2001,Nisoli:2013}.
In fact, frustrated magnetic models imply large degenerate classical ground states (GS) that are very sensitive to perturbations such as thermal or quantum fluctuations, spin-orbit interactions, spin-lattice couplings and impurities, all of which might be present in actual materials \cite{lacroix:2013,Diep:2013}. Novel unconventional phases such as valence bond solids and spin liquids can emerge from the effect of such purturbations on classical frustrated systems.
Moreover, the existence of artificial square ice \cite{Wang:2006,Wang:2007,Ke:2008} and the realization 
of quantum spin ice with Rydberg atoms \cite{Glaetzle:2014} demand a comprehensive understanding of
the associated models that are generic for such materials.

\begin{figure}
\includegraphics[width=\columnwidth]{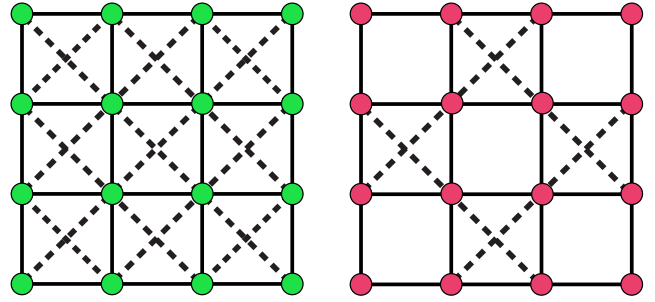}
\caption{(color online) $J_1-J_2$  model on (left) square lattice (SL) and (right) checkerboard lattice (CL). The solid and dashed lines 
are $J_1$ and $J_2$ bonds, respectively.}
\label{Fig1}
\end{figure}

Generally, a spin system is frustrated whenever
one cannot find a configuration of spins to fully satisfy the interacting
bonds between every pair of spins  \cite{Diep:2013,Moessner:2001}. 
For instance, a diagonal bond in addition to vertical and horizontal bonds construct a triangle, which makes frustration on the spins sitting on triangle corners of a square plaquette.
In this respect, spin 1/2 antiferromagnetic Ising models on the $J_1-J_2$ square and half depleted square, i.e. checkerboard, lattices are generic 2D frustrated magnets 
in which $J_1$ and $J_2$, the strength of nearest and next nearest neighbor interactions, respectively, compete with each other (see Fig.~\ref{Fig1}). These are prototype models that low dimensionality makes them an easier target for numerical/analytical approaches in contrast to 3D counterparts \cite{Siddharthan:1999,Tsui:1999,Gardner:1999,Melko:2001,Ruff:2005}. 
Accordingly, CL can be assumed as the 2D version of pyrochlore lattice of true spin ice materials \cite{Moessner:2004}. Here, we focus particularly on the role of quantum fluctuations on the ground state phase diagram of planar spin-ice, namely: CL and its low-energy effective theory on the square lattice.



In the case of Ising model on CL, quantum fluctuations introduced by both 
transverse magnetic field \cite{Moessner:2001prb,Moessner:2004} and
in-plane XY interactions \cite{Shannon:2004,Starykh:2005,Chan:2011,Bishop:2012} 
lift the classical degeneracy of the highly frustrated point $J_2=J_1$ toward a non-magnetic plaquette-VBS phase \cite{Moessner:2001prb,Moessner:2004,Shannon:2004,Sadrzadeh:2015} with broken translational symmetry, which shows two-fold degeneracy. The plaquette-VBS phase,
which is mediated by anharmonic quantum fluctuations as an order-by-disorder phenomenon \cite{Villain:1980,Bellier:2001,Henry:2014},
emerges from an exponentially
degenerate classical background, which can not be observed
within linear spin-wave theory \cite{Henry:2012,Sadrzadeh:2018} due to strong frustration.
In order to shed more light on the highly frustrated region, in the first part of our paper, we obtain the GS phase diagram of CL accurately by using a variational tree tensor-network (TTN) ansatz and compare it with previous studies. We use a novel unconstrained (gauge-free) TTN, 
generalized to CL, to approximate the ground state of the system with higher accuracy compared with previous isometric schemes \cite{Tagliacozzo:2009}. By computing local correlations and plaquette operator expectations, we find that a plaquette-VBS state is established at the low magnetic field around $J_2=J_1$ region of CL. Our results show that by increasing transverse magnetic field a second-order phase transition occurs at $\Gamma_{c}=0.28$ from the plaquette-VBS phase to paramagnetic phase.
The associated critical exponents are $\nu=1$ and $\gamma\simeq0.4$, where $\nu$
reveals the divergence of correlation length and $\gamma$ is an exponent, which governs the singularity in magnetic susceptibility. We do not observe any other critical point except the mentioned one, which rules out a canted N\'{e}el phase predicted by the Monte-Carlo study \cite{Henry:2014} at $J_2=J_1$. Our results of unconstrained TTN are in good agreement with the results of the cluster operator approach (COA) \cite{Sadrzadeh:2015}.

On the other hand, the $J_1-J_2$ TFI model on the square lattice shows
an emergent string-VBS phase at the fully frustrated point $J_2=0.5J_1$ 
\cite{Sadrzadeh:2016,Sadrzadeh:2018b}.
It can be expressed that quantum fluctuations by means of transverse field, lift the classical degeneracy toward a doubly degenerate VBS states along the horizontal or vertical directions of the square lattice called string-VBS phase. 
However, there is a possibility that such a phase can be extended to an intermediate region around the highly frustrated point $J_2=0.5J_1$, which is sandwiched between a N\'{e}el and striped antiferromagnetic states for small and large $J_2/J_1$, respectively \cite{Kalz:2009}. 
Accordingly, in the second part of our paper we consider a different strategy to clarify the quantum phase diagram of TFI model on the $J_1-J_2$ SL. We introduce a mapping from CL to SL, 
which leads to the GS phase diagram of the $J_1-J_2$ SL in terms of
the phase diagram of CL of the corresponding model.
In other words, we claim that the low-energy effective theory of frustrated TFI on CL is 
given by frustrated TFI on SL.
This mapping suggests a string-VBS order at the highly frustrated regime of SL,
which is in agreement with the results of COA \cite{Sadrzadeh:2016}.
It is worth mentioning that the TFI model could represent the large easy-axis anisotropic
limit of the antiferromagnetic $J_1-J_2$ Heisenberg model, where 
the true nature of a non-magnetic (VBS) phase is still under debate on SL \cite{Sachdev:1990,Zhitomirsky:1996,Zhang:2003,Capriotti:2003,Starykh:2004,Mambrini:2006,Haghshenas:2018,Haghshenas:2018May}
. Our results would be useful for further investigations in the latter model.

The paper is organized as follows. In the next section, we briefly introduce the model and different phases on CL. In Sec.~\ref{TTNANSATZ}, we inaugurate a numerical TTN technique to find accurately the quantum phase diagram of CL. Then, in Sec.~\ref{Map} we establish the mapping from CL to SL and derive the corresponding quantum phase diagram of SL. Finally, the paper is summarized and concluded in Sec.~\ref{Conclusion}. The details of introduced mapping have been presented in Appendix \ref{ap-a}.

\section{The model Hamiltonian}
\begin{figure}
\includegraphics[width=\columnwidth]{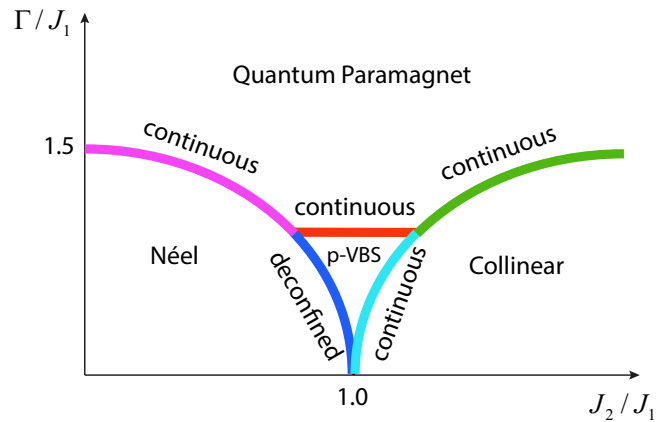}
\caption{(color online) A schematic phase diagram of the S=1/2 $J_1-J_2$ TFI model on CL \cite{Sadrzadeh:2015}, including the information on the type
of transitions between different phases obatined within TTN numerical simulation, namely: continious and deconfined phase transitions.}
\label{Fig2}
\end{figure}

The Hamiltonian of $J_1-J_2$ transverse field Ising model on CL is,
\begin{equation}
\label{eq1}
    \mathcal{H}=        J_1\displaystyle\sum_{\langle i,j \rangle}{S_i^zS_j^z}
                            +J_2\displaystyle\sum_{\langle\langle i,j \rangle\rangle}{S_i^zS_j^z}
                            -\Gamma\displaystyle\sum_{i}{S_i^x},~
\end{equation}
where $\langle i,j \rangle$ spans the nearest neighbor sites with $J_1$ coupling, 
$J_2>0$ is the diagonal coupling on crossed plaquettes, $\Gamma$ is the 
strength of transverse magnetic field and $S^{x,z}$ refer to {\it x} and {\it z} components of spin-1/2 operators 
on the vertices of the lattice (see Fig.~\ref{Fig1}). It consists of four different phases,
a N\'{e}el and collinear ordered phases close to the non-frustrated points $J_2/J_1=0$ and $J_2/J_1=2$ respectively, a quantum paramagnet phase at high fields and a plaquette-VBS phase for low magnetic fields $\Gamma\lesssim0.3$, a narrow region around the highly frustrated point $J_2=J_1$. 
The corresponding phase diagram is presented in Fig.~\ref{Fig2}, which has been obtained 
by COA approach \cite{Sadrzadeh:2015}.
In fact, it has been concluded that the exponential degeneracy of the classical ground state at the highly frustrated point, $J_2=J_1$, (known as square ice \cite{Henry:2012}) is lifted toward a unique quantum plaquette-VBS state that breaks translational symmetry of the lattice with two-fold degeneracy. 
It is a manifestation of order by disorder phenomena \cite{Villain:1980,Bellier:2001,Henry:2014}, which is induced by quantum fluctuations. 

In the next section, we use the unconstrained TTN approach to further confirm the quantum GS phase diagram of $J_1-J_2$ TFI model on CL. 
It has to be mentioned that the plaquette-VBS exists in a narrow region on the highly frustrated
regime, which requires to be investigated within high accurate numerical simulations.
In addition, we apply TTN to find critical points and critical exponents of the phase transitions from plaquette-VBS state to the N\'{e}el, collinear and paramagnet phases, which can classify the type of phase transitions.

%

\section{Unconstrained Tree Tensor Network ANSTAZ} 
\label{TTNANSATZ}
\begin{figure}
\includegraphics[width=\columnwidth]{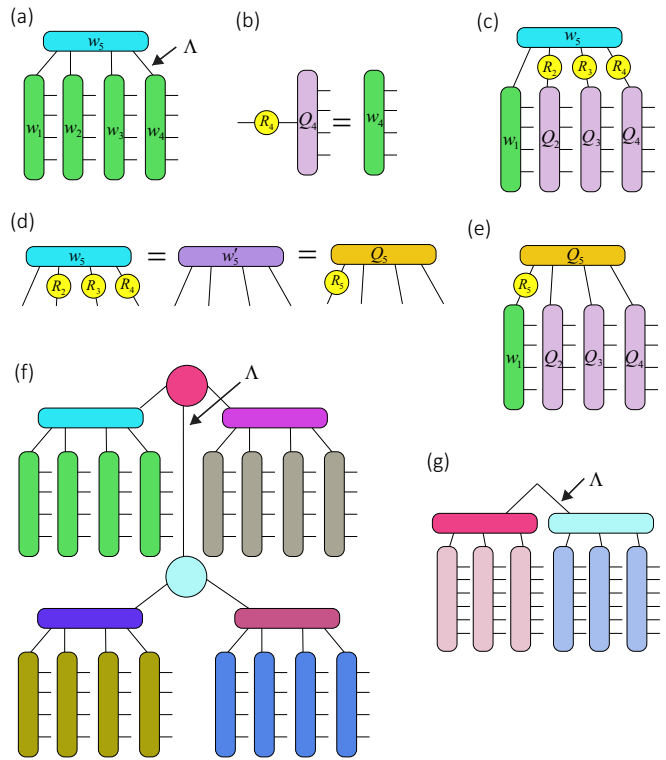}
\caption{(color online) Tensor-network representation of an unconstrained TTN state $|\Psi\rangle$ and its canonical form. (a) A TTN state for a $4 \times 4$ square lattice is represented by tensors $\{w_{i}\}$ connected by the so-called virtual bonds with dimension $\Lambda$ to form a tree-like geometrical graph. (b) A tensor-network representation of QR-decomposition applied to tensor $w_4=Q_{4}R_{4}$. One needs to fuse lower indices and then represents it in a matrix form to do decomposition. (c-e) The procedure to transform a general TTN state to a canonical normal form by using a sequence of QR-decomposition. The norm tensor is defined by removing tensor $w_1$ from tensor-network representation of $\langle \Psi_{w_1}|\Psi_{w_1} \rangle$, denoted by $\mathcal{N}$. A sequence of QR decomposition is used to make norm tensor identity $\mathcal{N}=I$: tensors $w_2, w_3, w_4$ are decomposed into QR forms, and (d) then tensors $R_{2}, R_{3}, R_{4}$ are absorbed into tensor $w_{5}$, followed by a QR-decomposition by fusing the virtual bonds (last ones) $w'_{5}=Q_5R_{5}$. (e) The canonical procedure is completed by absorbing $R_5$ into $w_1$, i.e. $w'_1=w_1R_5$. In this canonical form, one observes that $\langle \Psi|\Psi \rangle=\langle w'_1|I|w'_1\rangle$, i.e. the norm tensor is identity $\mathcal{N}=I$. The final optimum tensor $w'_1$ is obtained by solving $H_{eff}|w'_1\rangle=\lambda_{min} |w'_1\rangle$, where $H_{eff}$ are obtained by removing tensor $w'_1$ from $\langle \Psi_{w'_1}|H|\Psi_{w'_1} \rangle$.}
\label{Fig3}
\end{figure}

The TTN states provide a variational ansatz \cite{Shi:2006,Silvi:2010,Tagliacozzo:2009,Murg:2010,Gerster:2014} to simulate large 2D lattice sizes, beyond the possible sizes, 
which can be reached by exact diagonalization. We use an unconstrained TTN ansatz to variationally approximate the ground-state wave function of the TFI model (Eq.~\ref{eq1}) on the CL. The wave function is made of the local tensors $\{w_i\}$ connected to each other to form a tree-like graph as shown in Fig.~\ref{Fig3}-(a). The tensors $\{w_i\}$ effectively map a number of spins to an effective superspin by dimension $\Lambda$ at each layer, making a coarse-graining transformation---each tensor $w_i$ defines a projection from original (physical) Hilbert space onto the relevant subspace. That is the basic idea in the renormalization group (RG) methodology invented by Wilson and Kadanoff \cite{Efrati:2014}. Here, the goal is to use an efficient variational ansatz to minimize the ground-state energy with respect to tensors $\{w_i\}$, finding the best variational parameters (which grows like $\mathcal{O}(\Lambda^{3})$). In this paper, we use a recently introduced novel ansatz \cite{Gerster:2014} which, in contrast to traditional schemes, releases the internal gauge symmetry of the tensors (the isometry constraint) and provides a computationally stable and efficient algorithm with higher accuracy.

We shortly explain the unconstrained TTN variational ansatz generalized to two-dimensional lattices. The optimization method is performed by minimizing the energy with respect to a specific tensor $w_{i}$ (while holding fixed other tensors), i.e. 
\begin{align*}
&\min_{w_i} \{ \langle \Psi_{w_i} | H |\Psi_{w_i} \rangle   - \lambda \langle \Psi_{w_i}|\Psi_{w_i} \rangle  = \\
&\langle w_i | H_{eff} |w_i \rangle- \lambda \langle w_i|\mathcal{N}|w_i\rangle \},
\end{align*}
where the so-called norm tensor $\mathcal{N}$ and effective Hamiltonian $H_{eff}$ are obtained by removing tensor $w_i$ from the tensor-network representation of $\langle \Psi_{w_i}|\Psi_{w_i} \rangle$ and $\langle \Psi_{w_i}|H|\Psi_{w_i} \rangle$. The solution is given by solving a generalized eigenvalue problem $H_{eff}|w_i\rangle=\lambda_{min} \mathcal{N}|w_i\rangle$, which is a standard equation in linear algebra. The optimization procedure is then completed by using an iterative strategy: at each step, only one tensor is optimized while others hold fixed and then this task is repeated over all tensors till the variational energy does not change significantly. In practice, the norm tensor $\mathcal{N}$ causes instability in the algorithm, while the condition number (i.e. smallest singular value) would be too small. In order to avoid that, we need to use a `canonical normal form' \cite{Verstraete:2008} for the TTN state $|\Psi_{w_i} \rangle$ by making the norm tensor identity $\mathcal{N}=I$ (which is the best conditioning). The basic idea to do that is to use an appropriate gauge transformations similar to the case of matrix product states: it is obtained by using a sequence of QR-decomposition by fusing virtual bonds in a specific direction as shown in Fig.~\ref{Fig3}-(b-d). In this figure, we have explained how to use QR-decomposition to end up with a canonical form. Once we obtain that, we replace the tensor $w_i$ by solving standard eigenvalue problem $H_{eff}|w_i\rangle=\lambda_{min} |w_i \rangle$, which could be efficiently solved without suffering from bad conditioning.

The essential parameter $\Lambda$ controls the accuracy of TTN ansatz, as for $\Lambda \rightarrow \infty $ the TTN state faithfully represents the actual ground state of the system. The computational cost of the algorithm scales like $\mathcal{O}(\Lambda^{4})$ and $\mathcal{O}(\Lambda^{3})$ for running time and memory, respectively. 
In the present numerical TTN simulation, we consider clusters $4\times 4$, $6\times 6$ and $8\times 8$ with both periodic and open boundary conditions. We always do a finite-size analysis to study the behavior of the order parameters. A polynomial fit up to the fourth order is used 
to extrapolate the expectation values in $\Lambda \rightarrow \infty$ limit.
The largest bond dimension that we could afford is $\Lambda \sim 500$, so that error in the variational ground-state energy is at least of the order $10^{-4}$ (near the critical point, which is the less accurate case).

\subsection{$J_1-J_2$ TFI model on the checkerboard lattice: TTN results}
Before presenting the results, let us mention that the interesting and controversial part of 
TFI model on the CL is in the low magnetic field limit around the 
highly frustrated coupling $J_2=J_1$. This clarifies the reason that we concentrate on 
this region, while the other parts of the phase diagram are known by other 
methods without doubt \cite{Henry:2012,Sadrzadeh:2015}.
To obtain an accurate phase diagram for $J_1-J_2$ TFI model on the CL via TTN approach, 
we compute the first and second derivatives of the ground state energy by TTN simulation in two distinct directions on the phase diagram. 
Firstly, we trace the phase diagram along $\Gamma/J_1$ at fixed $J_2=J_1$ and then we 
consider another direction along $J_2/J_1$ at fixed magnetic field $\Gamma/J_1=cte$.

\begin{figure}
\centering
\includegraphics[width=0.9\columnwidth]{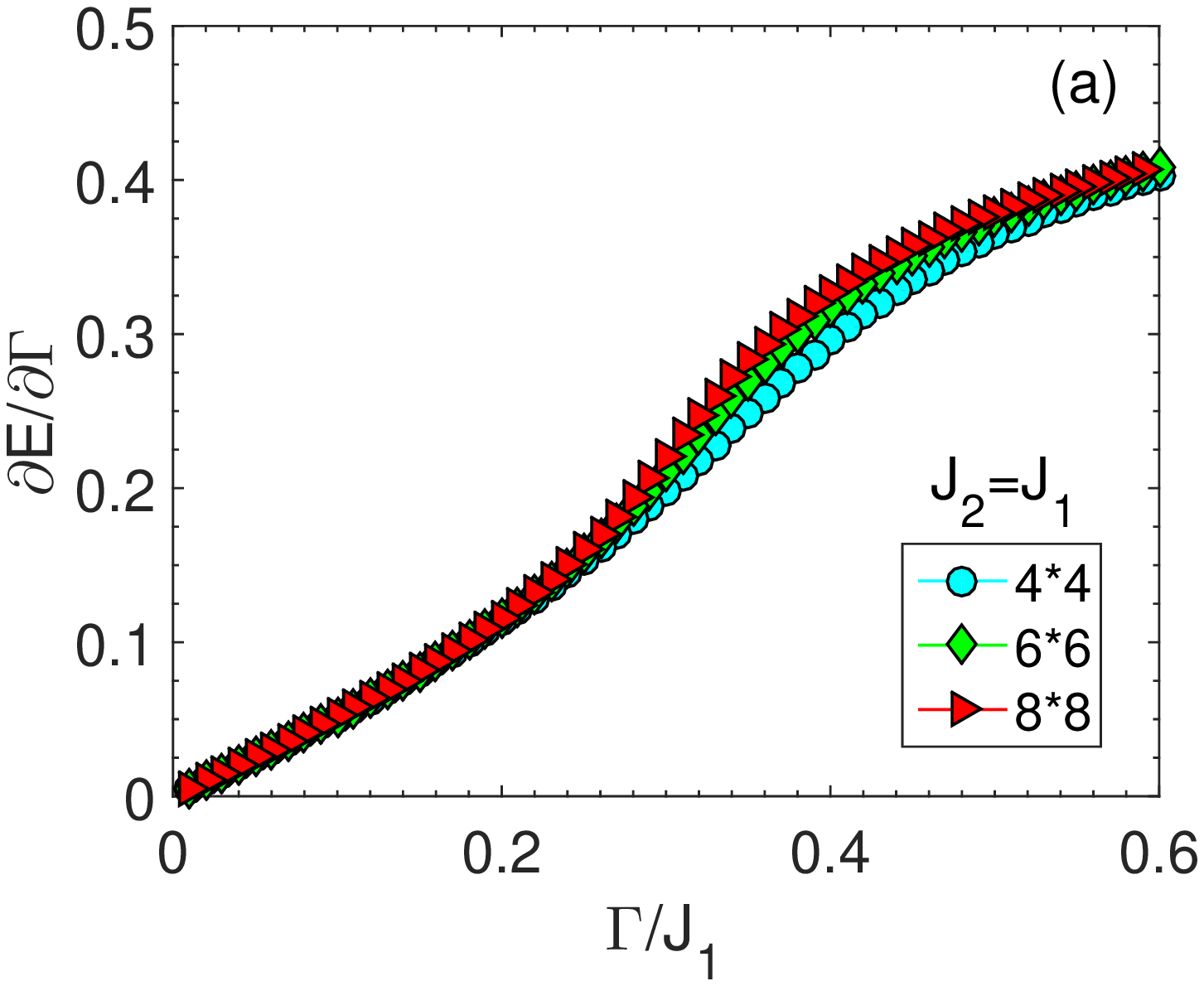}
\includegraphics[width=0.9\columnwidth]{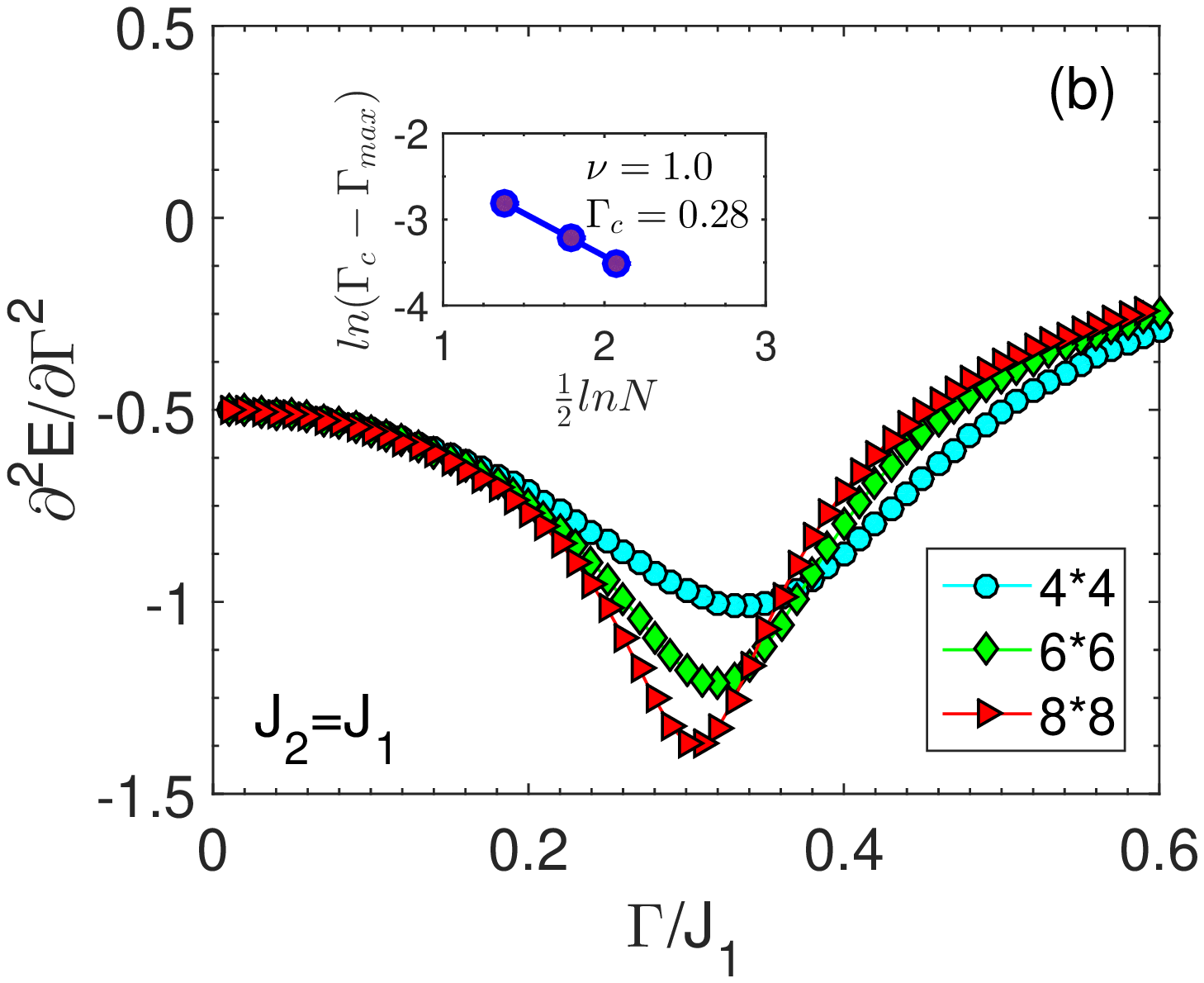}
\includegraphics[width=0.9\columnwidth]{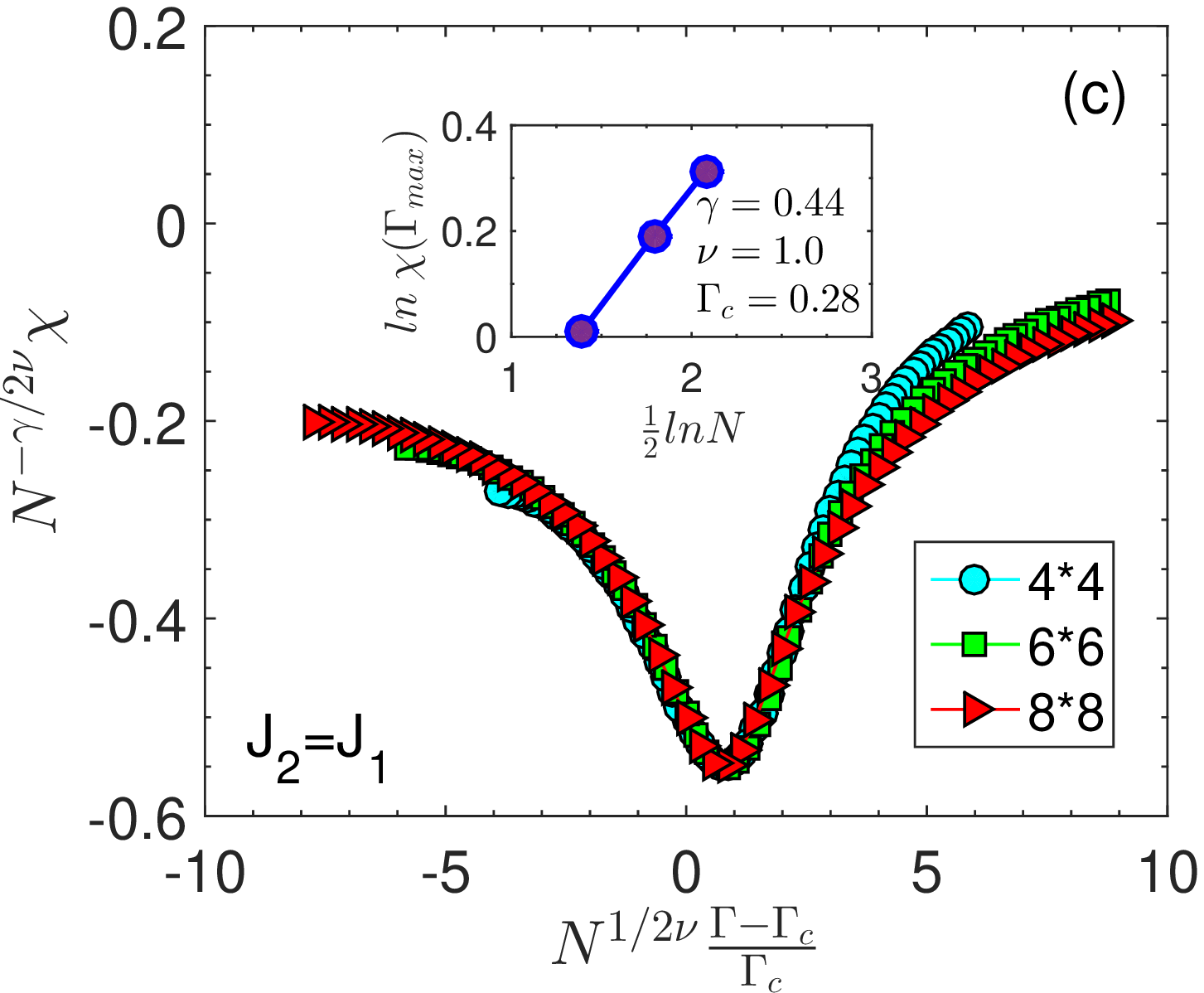}
\caption{(color online) (a) The first derivative of GS energy with respect to $\Gamma$ corresponding to the transverse magnetization obtained from TTN data for different system sizes. (b) 
The second derivative of GS energy with respect to $\Gamma$, which is the magnetic susceptibility obtained from TTN simulation for different lattice sizes. It shows only a sharp peak indicating a 
phase transition from the plaquette-VBS phase in low fields to the quantum paramagnet phase in high fields at $({\Gamma/J_1})_c = 0.28\pm 0.01$ with exponent $\nu=1.0 \pm 0.01$. (c) Data 
collapse of magnetic susceptibility obtained from TTN data, which shows the scale invariance of susceptibility governed by exponent $\gamma=0.44\pm 0.01$.}
\label{Fig4}
\end{figure}
\subsubsection{$J_2=J_1$}
According to the following equations, the first and second derivatives of ground state energy with respect to $\Gamma$ for the limit $J_2=J_1$ are equivalent to the transverse magnetization and magnetic susceptibility, respectively,
\begin{eqnarray}
\label{eq2}m_x&=&-\partial\langle \mathcal{H}\rangle/\partial \Gamma,\\
\chi &=& \partial m_x/\partial \Gamma=-\partial^2\mathcal{H}/\partial\Gamma^2.
\label{eq3}
\end{eqnarray}
Fig.~\ref{Fig4}-(a) and (b) show these quantities versus $\Gamma/J_1$ (at $J_2=J_1 $) obtained from TTN data for different lattice sizes. The transverse magnetization continuously reaches to its saturated value, which rules out any first order transition at this isotropic regime. However, we can see a peak on the magnetic susceptibility, which becomes sharper and stronger by increasing the lattice size, corresponding to a continious second order phase transition. We use finite-size scaling theory to evaluate the critical point and critical exponents for this transition \cite{Nishimori:2011}. 
The scaling behavior of $\chi$, which governs the singularity at the 
critical point is 
\begin{eqnarray}
 \label{eq4}|\Gamma_c - \Gamma_{max}| &\sim & N^{-1/2\nu},\\
 \chi(\Gamma_{max})&\sim & N^{\gamma/2\nu},
 \label{eq5}
\end{eqnarray}
where $\Gamma_c$ is the critical field in the infinite size,
$\Gamma_{max}$ is the position of extermum of finite-lattice susceptibility, 
$\nu$ is the correlation length exponent
i.e. $\xi \sim |\Gamma - \Gamma_c|^{-\nu}$ and $\gamma$ exhibits 
the trend of singularity in the magnetic susceptibility.

\begin{figure}
\includegraphics[width=0.6\columnwidth]{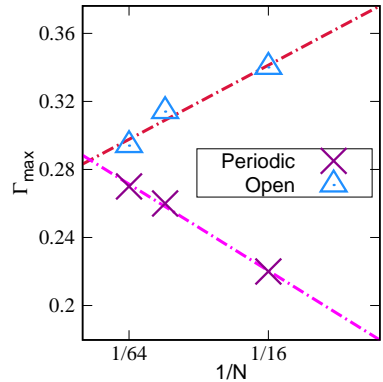}

\caption{(color online) The value of critical point versus inverse of lattice sizes.
Both periodic and open boundary conditions are presented, which are fitted by the scaling relation $\Gamma(N)=\Gamma_c(\infty)+\frac{a}{N}$. We obtain $\Gamma_c(\infty)=0.28\pm0.01$ and $0.285\pm0.01$ for open and periodic boundary conditions, respectively.
}
\label{Fig5}
\end{figure}

We found a good scaling of TTN data, which gives the critical field to be $\Gamma_c=0.28 \pm 0.01$ in the thermodynamic limit. Interestingly, Fig.~\ref{Fig5} confirms that both open and periodic boundary conditions lead to the same critical field $\Gamma_c\simeq0.28$. This critical point is also in a good accord with $\Gamma_c\simeq0.3$ obtained from COA results ~\cite{Sadrzadeh:2015}. The inset of Fig.~\ref{Fig4}-(b) shows the correlation length exponent obtained from finite-size scaling to be $\nu=1.0\pm 0.01$. Moreover, the scale-invariant behavior of magnetic susceptibility is shown in Fig.~\ref{Fig4}-(c) representing a good data collapse of different sizes with exponent $\gamma=0.44 \pm 0.01$. Furthermore, 
the presence of only one peak in magnetic susceptibility, assures that two distinct phases
exist at $J_2=J_1$, which are separated at $\Gamma_c$.
This single peak can be a signature for a quantum continuous phase transition from the plaquette-VBS phase at low fields to the quantum paramagnetic phase of high fields. The continuous nature of such transition is also confirmed by the broken lattice translational symmetry of the plaquette-VBS phase compared with symmetric quantum paramagnetic phase, as we expect from a Landau-Ginzburg paradigm. The TTN results presented on the large two-dimensional lattices 
$N = 4\times 4,\; 6\times 6$ and $8\times 8$ do not show any signature for another phase transition 
at $J_2=J_1$, which rules out the existence of a N\'{e}el order within $0.13\lesssim\Gamma\lesssim0.28$ that has been reported by Monte-Carlo simulation in Ref.~\cite{Henry:2014}.

\begin{figure}
\includegraphics[width=\columnwidth]{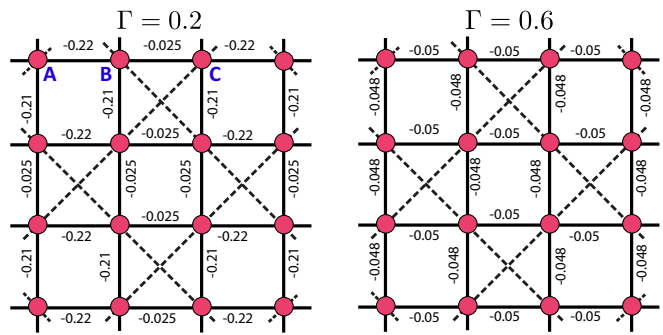}
\caption{(color online) Nearest-neighbor correlations, obtaind by TTN 
numerical simulation on the center of a $8\times8$ lattic at the $J_2=J_1$. 
Left: correlations at $\Gamma/J_1=0.2$, which shows the breaking of lattice translational  
symmetry corresponding to the plaquette-VBS phase. Right: correlations at $\Gamma/J_1=0.6$ corresponding to high-field regime of the quantum paramagentic phase, which preserves translational symmetry.}
\label{Fig6}
\end{figure}

In order to confirm the nature of the ground state at low fields, we calculate the nearest neighbor correlation function, $C_{NN}=\langle S_i^z S_j^z \rangle$, using TTN simulations on the $8\times8$ lattice at $J_2=J_1$. We obtained this correlation function for two different low and high values of transverse field $\Gamma$, shown in Fig.~\ref{Fig6}. Correlations for the low field regime depict a value close to the maximum value of N\'{e}el type ordering $C_{NN}^{max}=-0.25$ 
on the bonds of empty plaquettes with no corner sharing,
while correlations have very small values on the other plaquettes.
This is a clear signature of the plaquette formation as a VBS state, which breaks lattice translational symmetry leaving two-fold degeneracy. However, by increasing the magnetic field to the high field regime, we reach a quantum paramagnetic phase as it shows small correlations along vertical and horizontal directions of the lattice. 

\begin{figure}
\centering
\includegraphics[width=0.9\columnwidth]{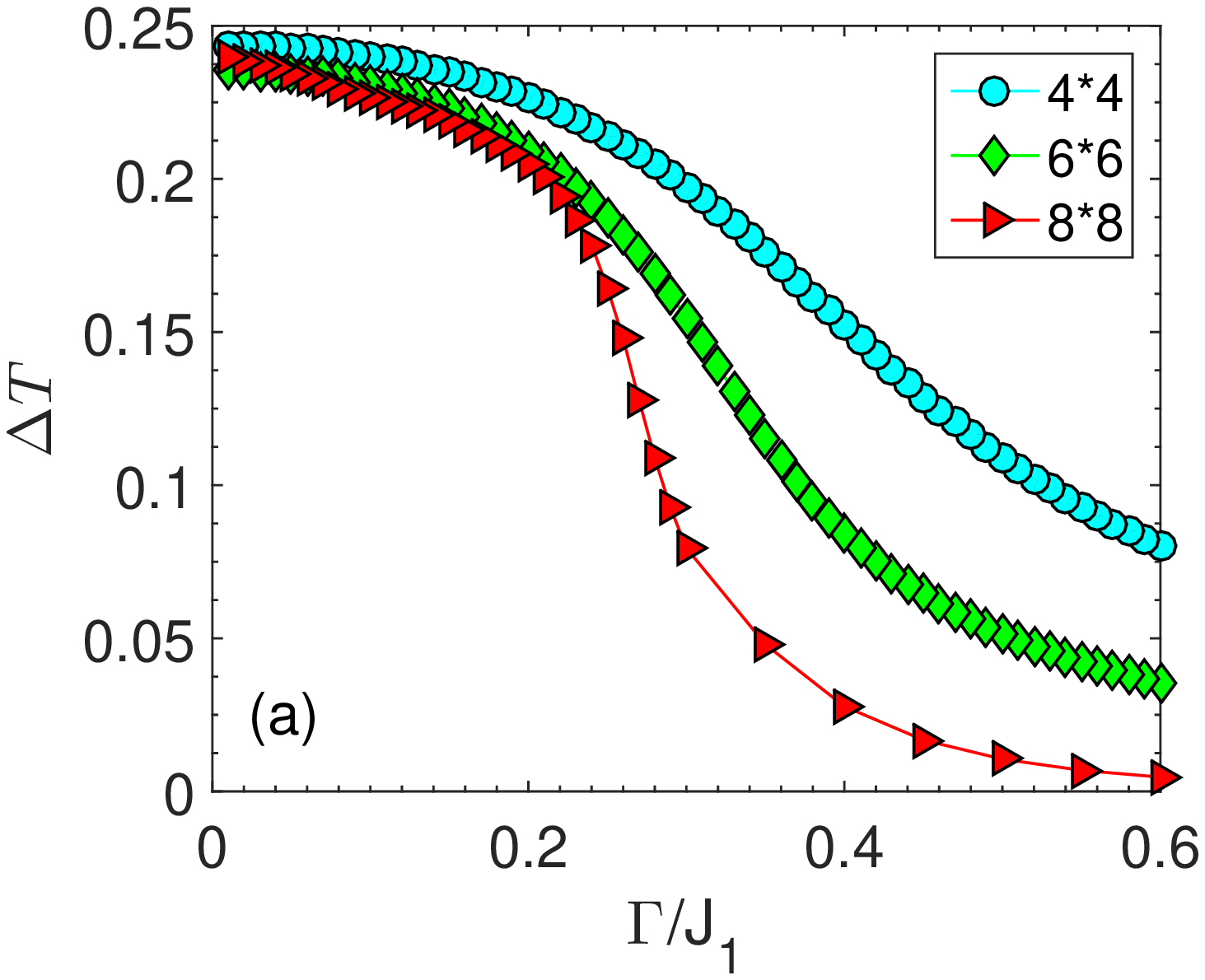}
\includegraphics[width=0.9\columnwidth]{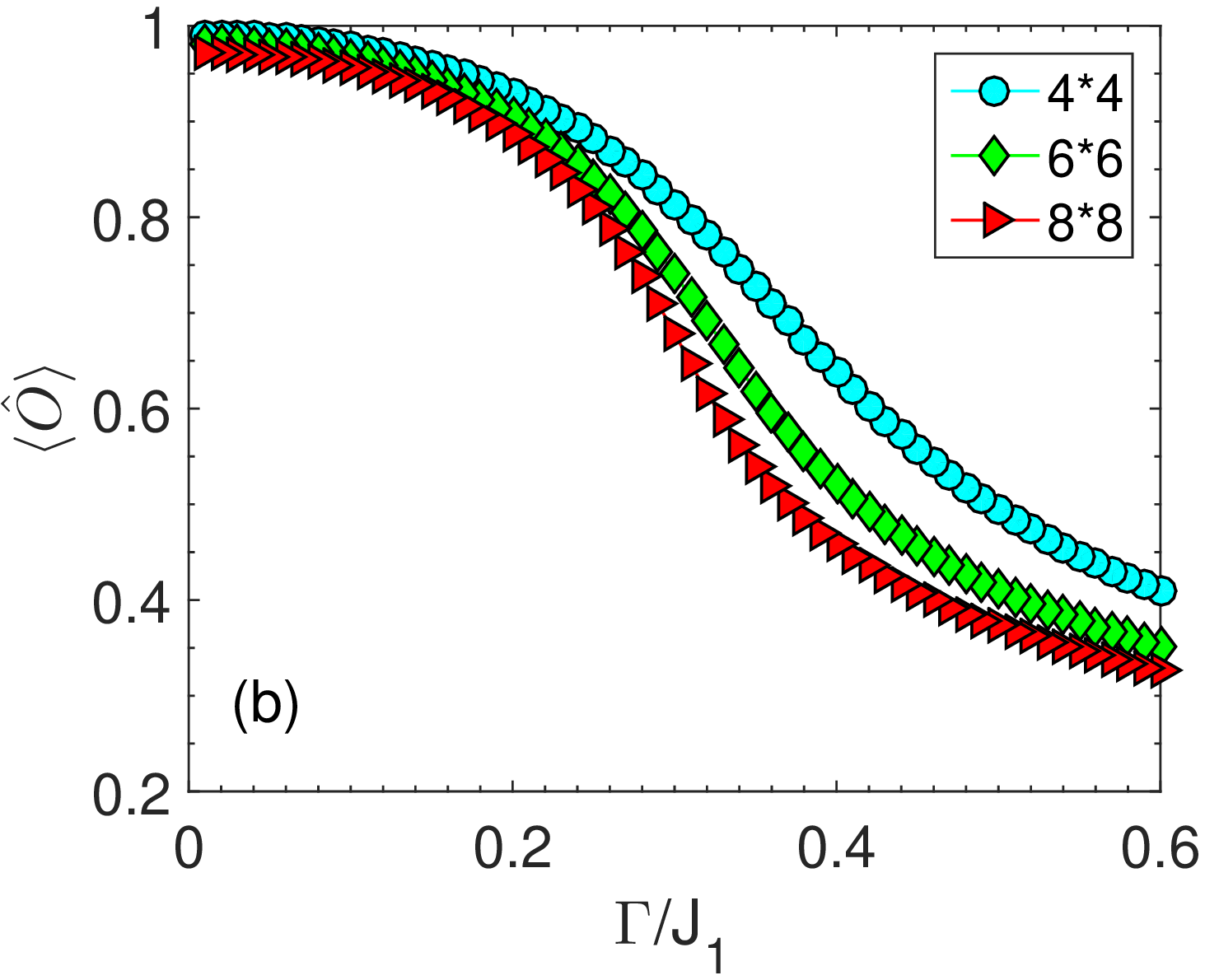}
\caption{(color online) (a) Expectation value of the translational order parameter $\Delta T$ and (b) the plaquette order parameter operator $\langle\hat{O}\rangle$ versus transverse magnetic field,
obtained by unconstrained TTN ansatz on different lattice sizes.}
\label{Fig7}
\end{figure}

Moreover, we plot in Fig.~\ref{Fig7}-(a), the translational order parameter, defined by 
\begin{equation}
 \Delta T=\langle S_A^z S_B^z \rangle -\langle S_B^z S_C^z \rangle, 
 \label{eq6}
\end{equation}
as a function of $\Gamma/J_1$ for different system sizes, where the sites A, B, and C are
shown in Fig.~\ref{Fig6}. It is observed that by increasing system size the translational order parameter rapidly decreases (extrapolates to zero in the infinite size limit) for $\Gamma > 0.3$ and tends to a finite value for $\Gamma < 0.3$ (lattice translational symmetry breaking), which is in agreement with the nature of the phases discussed above.

In addition, we support the plaquette-VBS nature of the ground state at low fields by calculating the ground state expectation value of resonating plaquette operator ($\hat{O}$) \cite{Henry:2014,Sadrzadeh:2015}. This operator is defined as 
\begin{equation}
 \hat{O}=\vert \varphi\rangle\langle {\bar \varphi}\vert+\vert {\bar \varphi}\rangle \langle \varphi \vert, 
 \label{eq7}
\end{equation}
where $\vert \varphi\rangle=\vert\uparrow\downarrow\uparrow\downarrow\rangle$ and $\vert {\bar \varphi} \rangle=\vert \downarrow\uparrow\downarrow\uparrow \rangle$ are two possible N\'{e}el configurations of a single plaquette. In fact, $\hat{O}$ defines a measure of resonating magnitude between $\vert \varphi\rangle$ and $\vert {\bar \varphi} \rangle$ on a plaquette. It is a suitable definition as it avoids formation of magnetic long range orders like N\'{e}el and collinear states on the whole lattice. Hence, the expectation value of $\hat{O}$ is very close to one for a resonating plaquette valence bond solid state, which has no magnetic order in z-direction. Fig.~\ref{Fig7}-(b) shows the expectation value of $\langle\hat{O}\rangle$ obtained by TTN simulation on different lattice sizes. It is evident that for $J_2=J_1$ and low fields, the value of $\langle\hat{O}\rangle$ is very close to unity which corresponds to the presence of a plaquette-VBS state.

\begin{figure}
\centering
\includegraphics[width=0.9\columnwidth]{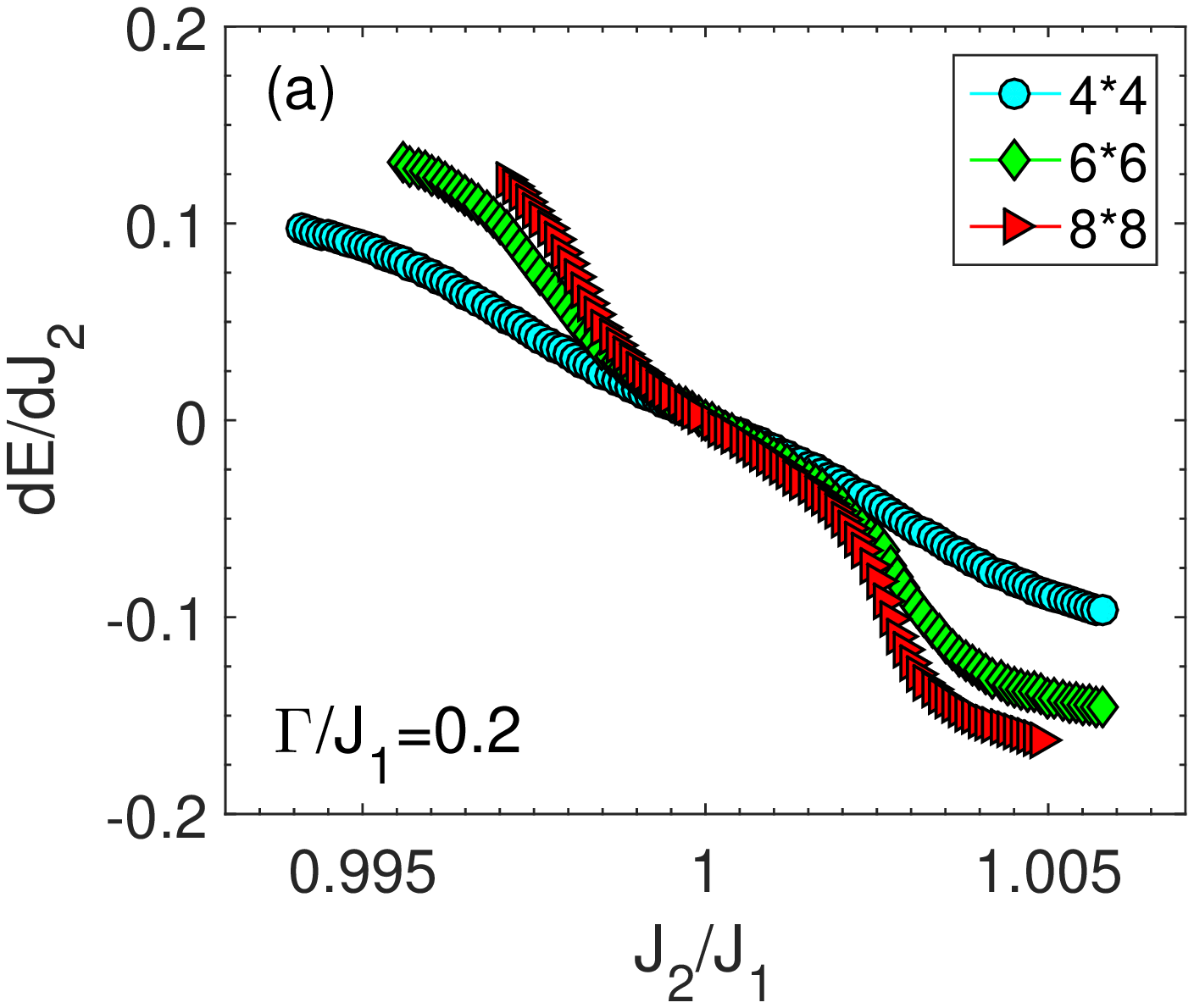}
\includegraphics[width=0.9\columnwidth]{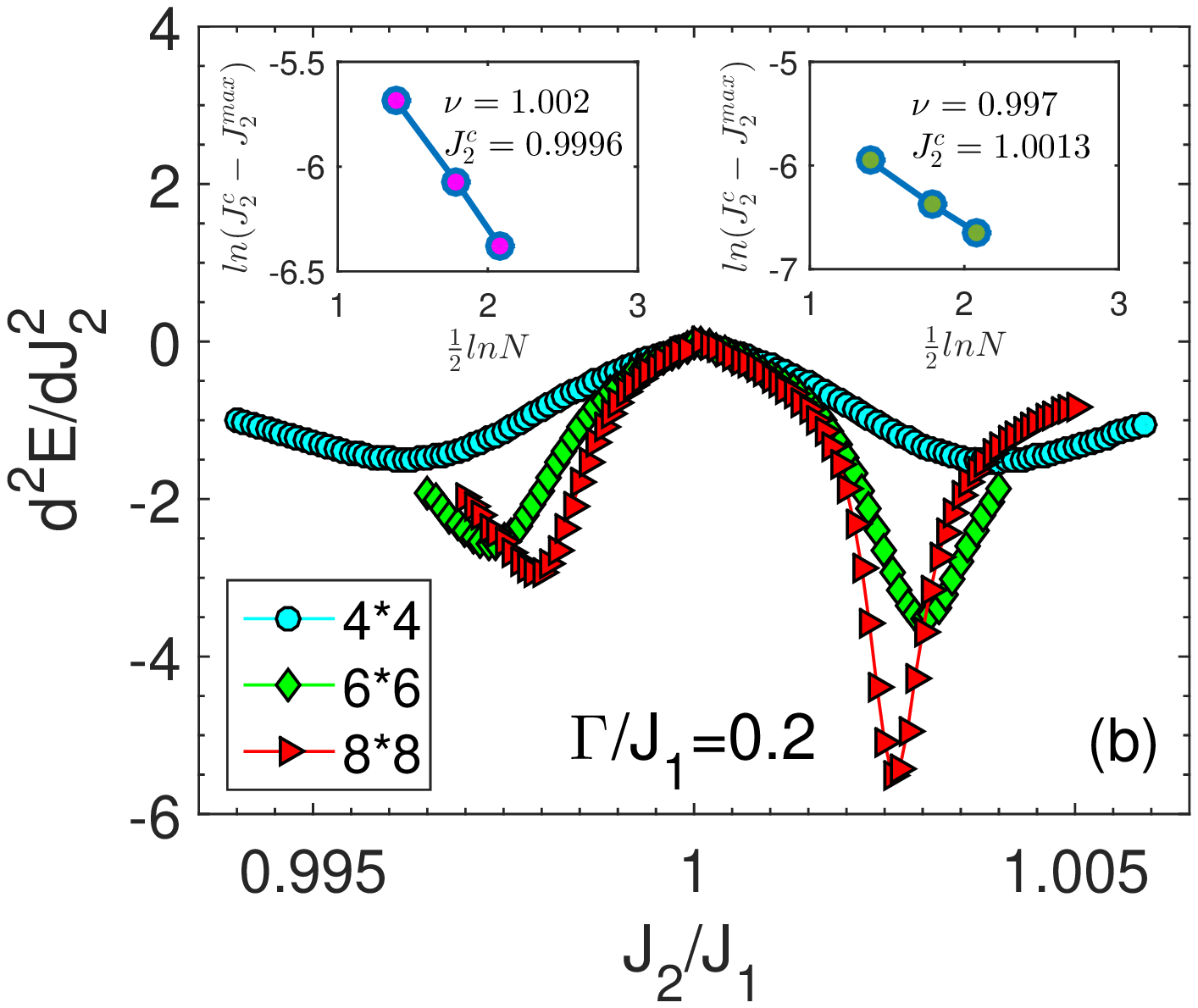}
\caption{(color online) (a) The first derivative of GS energy with respect to $J_2$ at $\Gamma/J_1=0.2$, obtained from TTN simulation for different system sizes, (b) The second derivative of GS energy with respect to $J_2$ at $\Gamma/J_1=0.2$, shows two sharp peaks indicating a phase transition from the N\'{e}el and collinear phases on both sides to the intermediate plaquette-VBS phase. The critical points occur at $({J_2/J_1})_c = 0.9996\pm 0.0001$ and $({J_2/J_1})_c = 1.0013\pm 0.0001$, respectively for N\'{e}el to plaquett-VBS and collinear to plaquette-VBS transitions, both with critical exponent $\nu\simeq1.0$. }
\label{Fig8}
\end{figure}

\begin{table}
\begin{tabular}{|c|c|c|c|c|}
\hline
$J_2=J_1$&$\Gamma=0.1$ &$\Gamma=0.2$ & $\Gamma=0.3$  & $\Gamma=0.4$\\
\hline
$E/N$ &-0.2525& -0.2607& -0.2770 & -0.3050 \\
\hline
$\langle\hat{o}\rangle$ &0.9563&0.8866& 0.6784 & 0.4579\\
\hline
\end{tabular}
\caption{ \label{table1}
Numerical results of ground state energy per site and plaquette order parameter obtained from TTN simulations on $8\times8$ CL at $J_2=J_1$ and open boundry condition.} 
\end{table}

\subsubsection{$\Gamma/J_1=0.2$}  
To elucidate the structure of phase diagram close to strong frustration, we fix 
the magnetic field in $\Gamma/J_1=0.2$ and trace the behvaior along $J_2/J_1$.
The first derivative of GS energy, according to relation $C^{(2)}=\langle S_{i}^{z}S_{j}^{z}\rangle_{\langle\langle i,j\rangle\rangle}=\partial{\langle \mathcal{H}\rangle}/\partial{J_{2}}$, is equivalent to the next-nearest neighbor spin-spin correlation. Fig.~\ref{Fig8}-(a) presents
$C^{(2)}$ versus $J_2/J_1$, which shows a change of sign at $J_2=J_1$.
However, the derivative of $C^{(2)}$---that is the second derivative of energy---represents two peaks as shown in Fig.~\ref{Fig8}-(b), which become sharper by increasing the lattice size. These peaks are interpreted as two critical points corresponding to two-phase transitions from the intermediate plaquette-VBS phase to the N\'{e}el and collinear phases on both sides of the phase diagram. The nature of quantum phase transition from the plaquette-VBS to N\'{e}el and collinear antiferromagnetic phases is an interesting feature of our results. The N\'{e}el and plaquette-VBS orders break different kind of symmetries, i.e. N\'{e}el order breaks a discrete $Z2$ symmetry while
plaquette-VBS breaks continuous translational symmetry. We might expect that the nature of this transition to be of the first order type, in terms of conventional Landau-Ginzburg theory. However, the first order transition is ruled out by no singular behavior in the first derivate
of GS energy as shown in Fig.~\ref{Fig8}-(a). Hence, we claim that the plaquette-VBS to N\'{e}el transition should be of a deconfined quantum continuous type according to the theory of deconfined quantum criticality~\cite{Senthil:2004}. 
The deconfined quantum critical point between N\'{e}el and plaquette-VBS phases occurs at $(J_2/J_1)_c=0.9996$, which is completely consistent with the COA data reporting $0.999$ \cite{Sadrzadeh:2015}. On the other hand, as seen from Fig.~\ref{Fig8}, the plaquette-VBS to collinear phase transition is also continuous. However, it would be a conventional second order phase transition, because both the plaquette-VBS and collinear phases break translational symmetry. The value of the latter critical point is $(J_2/J_1)_c=1.0013$, which is also in agreement with the value $1.001$ obtained by COA. The insets of Fig.~\ref{Fig8}-(b) depicts finite size scaling data which reports correlation length exponent to be $\nu\simeq1.0$ for both transition points.

As a summary, Tables.~\ref{table1} and \ref{table2} show some numerical results obtained by TTN simulation. Table.\ref{table1} represents numerical values of the ground state energy 
and plaquette order parameter at $J_2=J_1$ for different values of transverse field $\Gamma$. In Table.\ref{table2}, we tabulate the corresponding critical points and exponents obtained from finite-size scaling analysis on different parts of the phase diagram.

 \begin{table}
     \begin{ruledtabular}
\begin{tabular}{|c|c|c|c|c|c|c|}
\multicolumn{3}{|c|}{$J_2=J_1$} & \multicolumn{2}{c|}{$\Gamma=0.2$, \;$J_2<J_1$} & \multicolumn{2}{c|}{$\Gamma=0.2$, \;$J_2>J_1$}  \\
    
    \hline
$\Gamma_c$ & $\nu$  & $\gamma$ & $J_{2c}$ & $\nu$ & $J_{2c}$  & $\nu$ \\
\hline
0.28 & 1.0 & 0.44 & 0.9996 & 1.002 & 1.0013 & 0.997 \\
\end{tabular}

     \end{ruledtabular}
\caption{ \label{table2}
Numerical values of critical points and exponents resulting from the finite-size scaling analysis 
of TTN data for different regimes on the CL phase diagram.}
 \end{table}

\section{Map from the checkerboard lattice to the square lattice}
\label{Map}
\begin{figure}
\includegraphics[width=\columnwidth]{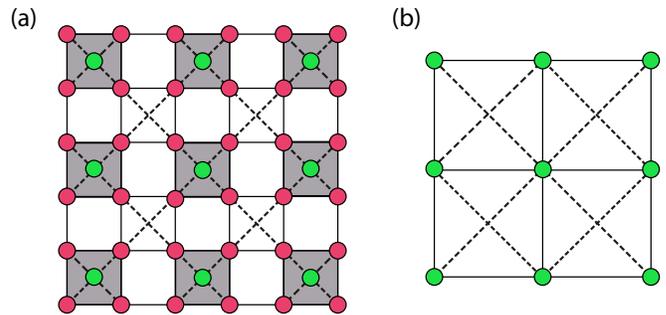}
\caption{(color online) Mapping from the CL to the square lattice. (a) Hatched crossed plaquettes of the CL form the unit cells of transformation. Solid and dashed lines are $J_1$ and $J_2$ bonds, respectively. Green bullets represent quasi spins, which are associated to each unit cell. (b) A square lattice constructed from quasi spins, by a lattice spacing twice as the original checkerboard one. Solid and dashed lines represent $J_1$ and $J_2$ bonds for the square lattice, respectively.}
\label{Fig9}
\end{figure}

\begin{figure}
\includegraphics[width=0.8\columnwidth]{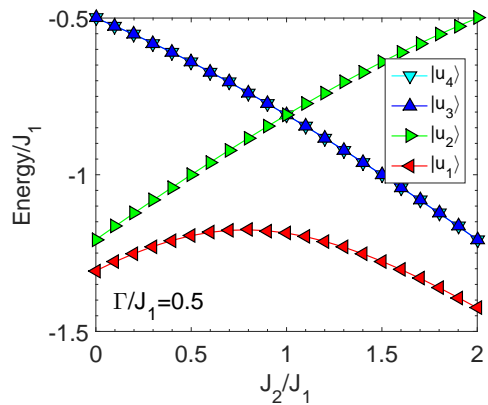}
\caption{(color online) The first four energy levels of a single crossed
plaquette spectrum versus $J_2/J_1$ in (an arbitrary) transverse field $\Gamma/J_1=0.5$.}
\label{Fig10}
\end{figure}

Here, we establish our map from CL to SL.
Let us consider non-corner sharing set of crossed plaquettes on CL, as unit cells of our transformation (see Fig.~\ref{Fig9}-(a)). According to Fig.~\ref{Fig9}-(a), we assign a quasi spin-half to each unit cell. These quasi spins form a new square lattice, whose lattice spacing is twice as the original lattice (see Fig.~\ref{Fig9}-(b)). 
Accordingly, the transverse field Ising Hamiltonian Eq.~\ref{eq1} can be rewritten in the form
\begin{eqnarray}
 H=&&H_0+H_{int}, \nonumber \\
 H_0=\sum_{I}{H_I}&,& \;\;\;\;H_{int}=\sum_{<IJ>}H_{IJ}, \label{eq8}
\end{eqnarray}
where $H_0$ is the sum on the Hamiltonians of unit cells and $H_{int}$ represents the interactions between unit cells. The Hamiltonian of a unit cell is diagonalized exactly, i.e. $J_1-J_2$ TFI model on a crossed plaquette with four spins. Fig.~\ref{Fig10} shows the first four energy levels of a unit cell, versus $J_2/J_1$ in an arbitrary transverse field $\Gamma$. 
For $J_2<J_1$, the first two eigenstates related to the lowest eigenenergies $\epsilon_1$ and $\epsilon_2$ are $\vert u_1\rangle$ and $\vert u_2\rangle$, respectively. 
These eigenstates are considered as the bases for a quasi-spin ($\hat{\tau}=1/2$) devoted to the unit cell.
Hence, we define $\vert u_1\rangle_I=\vert\tau_I^z=\uparrow\rangle$ and $\vert u_2\rangle_I=\vert\tau_I^z=\downarrow\rangle$. On the other hand, for $J_2>J_1$, the two eigenstates related to lowest eigenenergies are $\vert u_1\rangle$ and $\vert u_3\rangle$, where $\vert u_3\rangle$ is twofold degenerate, i.e. $\epsilon_3=\epsilon_4$. Therefore, for $J_2>J_1$, we consider two states $\vert u_1\rangle$ and $\vert u_2^{\prime}\rangle=\frac{1}{\sqrt{2}}(\vert u_3\rangle+\vert u_4\rangle)$ as $\vert\uparrow\rangle$ and $\vert\downarrow\rangle$ quasi-spins, respectively. 

In the next step, we define projecting operators onto the subspace spanned by the low-energy
sector of unit cells.
In fact, the terminology of effective theory, which describes the low-energy behavior of a model 
is always accompanied by the reduction in the Hilbert space. We define two projecting operators 
$P_{I}$ and $P_{I}^{{\prime}}$ of unit cell labeled by $I$, for $J_2<J_1$ and $J_2>J_1$, respectively. They read as, 
\begin{eqnarray}
\label{eq9}P_{I}&=&|u_{1}\rangle_{I I}\langle u_1|+|u_2\rangle_{I I}\langle u_2|,\\
\label{eq10}P_{I}^{\prime}&=&|u_1\rangle_{I I}\langle u_{1}|+|u_{2}^{\prime}\rangle_{I I}\langle u_{2}^{\prime}|.
\end{eqnarray}
These local operators act as Identity operator on other unit cells. Therefore, the projecting operator for the whole lattice is defined as $P=\bigotimes_I{P_I}$ and $P^{\prime}=\bigotimes_I{P_I^{\prime}}$. Hence, the effective Hamiltonian in truncated subspace will be obtained from the following relations,
\begin{eqnarray}
\label{eq11}\mathcal{H}_{eff}=P(\mathcal{H}_0+\mathcal{H}_{int})P,\;\;\;(J_2<J_1)\\
\label{eq12}\mathcal{H}_{eff}=P^{\prime}(\mathcal{H}_0+\mathcal{H}_{int})P^{\prime},\;\;\;(J_2>J_1).
\end{eqnarray}
The explicit form of $\mathcal{H}_0$ and $\mathcal{H}_{int}$ in terms of original spin operators are given in Appendix~\ref{ap-a}. 

The original Hamiltonian is renormalized in truncated subspace according to Eqs. \ref{eq11} and \ref{eq12}, which leads to the effective Hamiltonian as follows, 
\begin{eqnarray}
\nonumber J_2&<&J_1:\\
\label{eq13}\nonumber\mathcal{H}^{eff}&=&- 2 \alpha^2 J_1\displaystyle\sum_{\langle I,J \rangle}{\tau_I^x\tau_J^x}
                            +  \alpha^2 J_2\displaystyle\sum_{\langle\langle I,J \rangle\rangle}{\tau_I^x\tau_J^x}\\
                             &&-(\epsilon_2-\epsilon_1)\displaystyle\sum_{I}{\tau_I^z},
                            \\                        
\nonumber J_2&>&J_1:\\
\nonumber\mathcal{H}^{eff}&=& -2 \alpha^{{\prime}^2} J_1\displaystyle\sum_{\langle I,J \rangle_v}{\tau_I^x\tau_J^x}+ 2 \alpha^{{\prime}^2} J_1\displaystyle\sum_{\langle I,J \rangle_h}       {\tau_I^x\tau_J^x}\\
 &&-  \alpha^{{\prime}^2} J_2\displaystyle\sum_{\langle\langle I,J \rangle\rangle}{\tau_I^x\tau_J^x}
                            -(\epsilon_3-\epsilon_1)\displaystyle\sum_{I}{\tau_I^z}\label{eq14},                                    
\end{eqnarray}
where, $\langle I,J \rangle_h$ and $\langle I,J \rangle_v$ run over horizontal and vertical nearest neighbor bonds on the effective square lattice. The coefficients $\alpha$ and $\alpha^\prime$ are functions of $J_1$, $J_2$ and $\Gamma$ (see Appendix). 
Let us make a $\pi$-rotation around z-axis on the spins of one of the sublattices of the 
bi-partite square lattice defined in Eq.~\ref{eq13}, which contracts the 
the minus sign in the first term. Similarly, a $\pi$-rotation around z-axis on the spins sitting on
even (or odd) labeled horizontal lines change the minus signs of the first and third 
terms of Eq.~\ref{eq14}. Hence, all Ising terms ($\tau_I^x\tau_J^x$) in Eqs.~\ref{eq13}, \ref{eq14} have positive couplings. Now, it is clear from the sign of nearest and next-nearest neighbor interactions of the effective Hamiltonian, that there is a N\'{e}el and striped order for $J_2\ll J_1$ and $J_2\gg J_1$ limits, respectively. They correspond to well known classical magnetic ordered phases of the Ising model on the $J_1-J_2$ square lattice~\cite{MoranLopez:1993}. Hence, we can merge the
two effective Hamiltonians \ref{eq13} and \ref{eq14} and 
write a unified effective Hamiltonian in terms of the renormalized parameters
$\tilde{J}_1-\tilde{J}_2$ that is a transverse field Ising model on the effective square lattice, 
\begin{eqnarray}
\mathcal{H}^{eff}&=& \tilde{J_1}\displaystyle\sum_{\langle I,J \rangle}{\tau_I^x\tau_J^x}
                            +  \tilde{J_2}\displaystyle\sum_{\langle\langle I,J \rangle\rangle}{\tau_I^x\tau_J^x}-\tilde{\Gamma}\displaystyle\sum_{I}{\tau_I^z},
\label{eq15}                                             
\end{eqnarray}
where, 
\begin{eqnarray}
\label{eq16}\frac{\tilde{J_2}}{\tilde{J_1}}&=&\frac{1}{2}\frac{J_2}{J_1}, \nonumber \\
\frac{\tilde{\Gamma}}{\tilde{J_1}}&=&\frac{\epsilon_2-\epsilon_1}{2 \alpha^2J_1},\;\;\; (J_2<J_1),  \\
\frac{\tilde{\Gamma}}{\tilde{J_1}}&=&\frac{\epsilon_3-\epsilon_1}{2 \alpha^{{\prime}^2}J_1},\;\;\; (J_2>J_1). \nonumber
\end{eqnarray}
According to Eq.~\ref{eq15}, the low-energy effective theory of TFI model on CL 
is provided with the same model on a square lattice with renormalized parameters given in Eq.~\ref{eq16}.
The effective Hamiltonian clearly shows that at the zero field limit, the critical point $J_2=J_1$ of CL is mapped to the critical point $\tilde{J_2}=0.5\tilde{J_1}$ of SL (see Eq.~\ref{eq16}). Hence, the critical phase boundaries of $\tilde{J}_1-\tilde{J}_2$ TFI model on SL can be achieved from the critical phase boundaries of the $J_1-J_2$ TFI model on CL.

\subsection{GS phase diagram of $J_1-J_2$ TFI model on the square lattice}
\label{Result}
\begin{figure}
\includegraphics[width=\columnwidth]{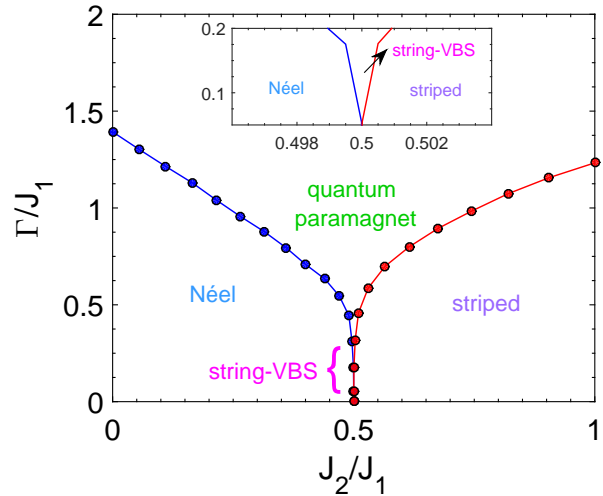}
\caption{(color online) Quantum GS phase diagram of the $J_1-J_2$ TFI model on square lattice obtained from the phase diagram of the CL \cite{Sadrzadeh:2015} using the introduced effective theory. The inset indicates an opening of a narrow region of string-VBS phase, which fills the space between the N\'{e}el and striped phases around $J_2/J_1=0.5$ for low fields.}
\label{Fig11}
\end{figure}
We implement the mapping established in the previous section and apply it to the GS phase diagram
of TFI model on CL--- which has been obtained by COA, \cite{Sadrzadeh:2015}--- to get the GS phase diagram of $J_1-J_2$ TFI model on SL. 
To this end, we insert the location of critical boundaries of the CL phase diagram in Eqs.~\ref{eq16} to obtain the corresponding critical boundaries of the SL phase diagram. The outcome of this map is shown in Fig.~\ref{Fig11}. 
For instance, the critical point $\Gamma_c/J_1=0.3$ at $J_2=J_1$ on CL
is mapped to $\Gamma_c/J_1=0.32$ at $J_2=0.5J_1$ on SL.
This result is consistent with the result $\Gamma_c/J_1=0.51$ obtained from TTN and COA data on the square lattice~\cite{Sadrzadeh:2016}. Moreover, Fig.~\ref{Fig11} demonstrates 
the presence of a narrow region around $J_2=0.5J_1$ at low fields, exactly the same as what appeared in the phase diagram of CL around the highly frustrated point $J_2=J_1$ at low fields, like Fig.~\ref{Fig2}. Hence, it can be deduced that quantum fluctuations of the weak transverse magnetic field induce a novel quantum state from the highly degenerate classical GS of SL at $J_2=0.5J_1$, before reaching to the quantum paramagnet phase at high fields.  
\begin{figure}
\includegraphics[width=\columnwidth]{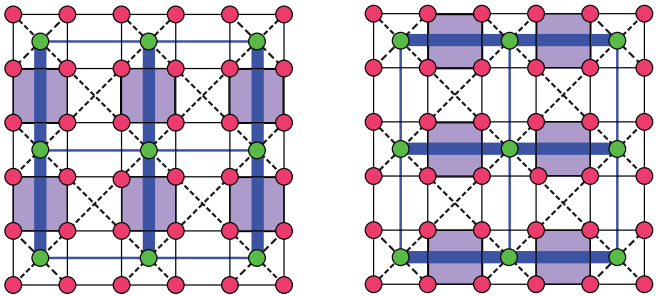}
\caption{(color online) Plaquette-VBS phase of CL with broken translational symmetry with two-fold degeneracy, which is mapped to the string-VBS phase of square lattice with broken rotational symmetry and two-fold degeneracy.}
\label{Fig12}
\end{figure}

One of the smart features of the introduced mapping is to determine the structure of the novel state according to the plaquette-VBS state on CL. Let us suppose that
the CL is in the plaquette-VBS phase as shown by the color
plaquettes in Fig.~\ref{Fig12}. 
In fact, each color plaquette is surrounded by two close sites on the effective square lattice. Therefore, whenever color plaquettes of CL resonate between two possible N\'{e}el states, which comes from the nature of plaquette-VBS phase, then they bring about a resonant situation on a set of sites on the effective square lattice resembling
the string formation. Moreover, as the plaquette-VBS state of CL breaks the translational symmetry of the lattice bearing two-fold degeneracy, the emergence of strings on the effective SL could be either in vertical or horizontal directions, breaking the rotational symmetry of the lattice, which manifests the two-fold degeneracy of string formations. This is in agreement with our earlier results in Ref.~\cite{Sadrzadeh:2016},
which states that the highly degenerate classical ground state of $J_1-J_2$ TFI model on SL at $J_2=0.5J_1$ goes to a unique string-VBS phase, when taking into account quantum fluctuations.
This justifies the mapping procedure introduced here.

\section{Summary and Conclusions}
\label{Conclusion}

Transverse field Ising model on two-dimensional checkerboard/square lattice
would be a generic Hamiltonian to
represent uni-axial magnets driven by quantum fluctuations. 
It includes planar spin ice \cite{Moessner:2004}, 
artificial square ice \cite{Wang:2006,Wang:2007,Ke:2008} and even the realization 
of quantum spin ice with Rydberg atoms \cite{Glaetzle:2014} that offer the emegence of 
novel phases. We have investigated the phase diagram of the $J_1-J_2$ TFI model on
checkerboard lattice by an improved tree tensor-network algorithm. We developed an unconstrained (gauge-free) tree tensor-network ansatz, adapted to two-dimensional systems up to the lattice size $8 \times 8$, by relaxing isometry constraint. At the highly frustrated point $J_2=J_1$, we confirm a plaquette-VBS phase at low fields, separated from a paramagnet phase at $\Gamma_{c}\sim 0.28$. Utilizing finite-size scaling analysis on $N=4\times 4,\; 6\times 6$ and $8\times 8$ lattices, we obtain the associated critical exponents to be $\nu \simeq 1$ and $\gamma \simeq 0.44$. We did not observe a signature of a canted N\'{e}el phase predicted by the Monte-Carlo study \cite{Henry:2014}, which is in agreement with previous results based on cluster operator approach \cite{Sadrzadeh:2015}. In addition, we found the nature and associated critical exponents of the quantum phase transitions from the plaquette-VBS phase to the adjacent N\'{e}el and collinear antiferromagnetic phases and also to the quantum paramagnetic phase of high fields, summarized in table-\ref{table2}. It is shown that all transitions are of the second-order type except the transition from N\'{e}el to plaquette-VBS,
which is of deconfined type, where the first derivative of ground-state energy indicates no singularity. The schematic structure of the phase diagram is given in Fig.~\ref{Fig2}.

Our study justifies the importance of unconstrained TTN ansatz as a promising numerical tool to address such highly frustrated systems, where quantum Monte Carlo simulation fails due to the known sign problem for reaching ground state properties. Furthermore, we have developed a mapping analysis to obtain quantum ground state phase diagram of the $J_1-J_2$ TFI model on square lattice from the phase diagram of the $J_1-J_2$ TFI model on checkerboard lattice. An important outcome of our mapping is to clarify the VBS nature of the intermediate phase of square-lattice phase diagram at low fields around the highly frustrated point $J_2=0.5J_1$. In fact, we showed that the plaquette-VBS phase of the checkerboard lattice is mapped to the string-VBS phase of sqaure lattice at the highly frustrated point $J_2=0.5J_1$, completely in agreement with the previous results of $J_1-J_2$ TFI model on square lattice by cluster operator approach, which describes such VBS ordering \cite{Sadrzadeh:2016}. Briefly, we claim that the low-energy effective theory of $J_1-J_2$ TFI model on checkerboard is given by the same model on square lattice with renormalized parameters.

\section{Acknowledgements}
A.L. would like to thank the Sharif University of Technology for financial support under grant No. G960208. R.H. was supported by the Department of Energy, Office of Basic Energy Sciences, Division of Materials Sciences
and Engineering, under Contract No. DE-AC02-76SF00515 through SLAC National Accelerator Laboratory. We have used \emph{Uni10} \cite{Kao:2015} library to build the TTN ansatz.

\appendix
\section{Mapping from the checkerboard lattice to the square lattice} 
\label{ap-a}

\begin{figure}[ht!]
\includegraphics[width=0.6\columnwidth]{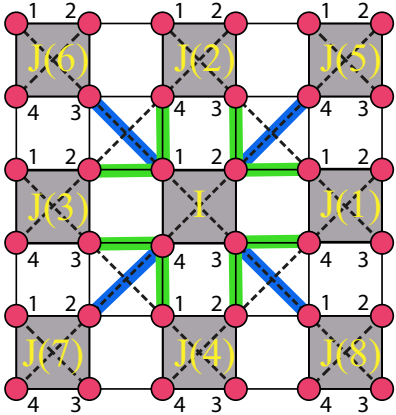}
\caption{(color online) The CL: each isolated plaquette I interacts with eight neigboring plaquettes. 
The green and blue lines correspond respectively to $J_1$ and $J_2$ interactions of plaquette I with its neighbors. }
\label{appendixa}
\end{figure}

The details of mapping procedure is presented here. As we explained in the text, if we divide the CL into non-corner-sharing crossed plaquettes, the transverse field Ising Hamiltonian can be rewritten in the form $\mathcal{H}=\sum_{I}{H_I}+\sum_{<IJ>}H_{IJ}$, where $H_I$ is the Hamiltonian on a single plaquette and $H_{IJ}$ defines the interaction Hamiltonian between single plaquettes. Fig.\ref{appendixa} depicts a typical single plaquette $I$ sorrounded by eight independent plaquettes $J$ interacting with it. According to site labeling of Fig.\ref{appendixa} we arrive at the following expression for $H_I$ and $H_{IJ}$,

\begin{eqnarray}
\label{eqap1}
H_I &&=  J_{1} (s_{1,I}^{z} s_{2,I}^{z} +s_{2,I}^{z} s_{3,I}^{z}+s_{3,I}^{z} s_{4,I}^{z}+s_{4,I}^{z} s_{1,I}^{z} ) \\
\nonumber &&+J_2 (s_{1,I}^{z} s_{3,I}^{z}+s_{2,I}^{z} s_{4,I}^{z} )-\Gamma (s_{1,I}^{x}+s_{2,I}^{x}+s_{3,I}^{x}+s_{4,I}^{x}),\\
\nonumber {H}_{IJ}&&={J_1(s_{2,I}^z s_{1,J(1)}^z+s_{3,I}^z s_{4,J(1)}^z)}+{J_2 (s_{2,I}^z s_{4,J(5)}^z)}+\\
\nonumber && {J_1 (s_{1,I}^z s_{4,J(2)}^z+s_{2,I}^z s_{3,J(2)}^z)}+{J_2 (s_{1,I}^z s_{3,J(6)}^z)}+\\
\nonumber && {J_1(s_{1,I}^z s_{2,J(3)}^z+s_{4,I}^z s_{3,J(3)}^z)}+{J_2(s_{4,I}^z s_{2,J(7)}^z)}+\\
&& J_{1} (s_{4,I}^z s_{1,J(4)}^z+s_{3,I}^z s_{2,J(4)}^z)+J_2(s_{3,I}^z s_{1,J(8)}^z).
\label{eqap2}
\end{eqnarray}
Let us consider the case $J_1>J_2$,  we consider the first two eigenstates $\vert u_1\rangle$ and $\vert u_2\rangle$ of $H_I$ --corresponding to the first two energy levels of it-- as two components of new quasi-spins assigned to each single plaquette. Then, we define the projecting operator $P_0$ as $P_0=|u_1\rangle\langle u_1|+|u_2\rangle\langle u_2|$ to renormalize original spin operators in the truncated subspace according to the following equations,
\begin{eqnarray} 
\nonumber P_0 s_1^z P_0&=&P_0 s_3^z P_0=\alpha \tau_I^x,\\
\nonumber P_0 s_2^z P_0&=&P_0 s_4^z P_0=-\alpha \tau_I^x,\\
 \nonumber P_0 s_1^x P_0&=&P_0 s_2^x P_0=P_0 s_3^x P_0=P_0 s_4^x P_0=(\beta-\gamma)\tau_I^z\\
\label{eqap3}
\end{eqnarray}
where, $\alpha=4A_2 B_1+2A_4 B_2$, $\beta=2A_2(A_1+2A_3+A_4)$ and $\gamma=2B_1B_2$ in which the coefficients $A$, $B$ are given by the matrix elements of eigenvectors $\vert u_1\rangle$ and $\vert u_2\rangle$,
\begin{eqnarray}
\vert u_1\rangle =
\begin{pmatrix}
A_1\\
A_2\\
A_2\\
A_3\\
A_2\\
A_4\\
A_3\\
A_2\\
A_2\\
A_3\\
A_4\\
A_2\\
A_3\\
A_2\\
A_2\\
A_1
\end{pmatrix}\;\;,\;\;
\vert u_2\rangle =
\begin{pmatrix}
0\\
B_1\\
-B_1\\
0\\
B_1\\
B_2\\
0\\
B_1\\
-B_1\\
0\\
-B_2\\
-B_1\\
0\\
B_1\\
-B_1\\
0
\end{pmatrix}.
\label{eqap4}
\end{eqnarray}

These matrix elements are functions of $J_1,J_2$ and $\Gamma$, which are lengthy and complicated expressions. The simplest one is $B_1$, which has the following form,
\begin{eqnarray}
B_1&=&-\frac{2}{\sqrt{\frac{2 \left(\sqrt{16 \Gamma ^2+({J_2}-2 {J_1})^2}+2 {J_1}-{J_2}\right)^2}{\Gamma ^2}+32}}.
\label{eqap5}
\end{eqnarray}

Now, we rewrite the Hamiltonians $H_I$ and $H_{IJ}$ of Eq.\ref{eqap1} and Eq.\ref{eqap2} in terms of new quasi-spins and finally obtain the effective Hamiltonian,
\begin{eqnarray}
\nonumber J_2&<&J_1:\\
\nonumber\\
\nonumber\mathcal{H}^{eff}&=&- 2 \alpha^2 J_1\displaystyle\sum_{\langle I,J \rangle}{\tau_I^x\tau_J^x}
                            +  \alpha^2 J_2\displaystyle\sum_{\langle\langle I,J \rangle\rangle}{\tau_I^x\tau_J^x}\\
                            \nonumber &&-(\epsilon_2-\epsilon_1)\displaystyle\sum_{I}{\tau_I^z},\\
\label{eqap6}
\end{eqnarray}
where $\epsilon_1$ and $\epsilon_2$ are eigenenergies of a single plaquette, corresponding to eigenvectors $\vert u_1\rangle$ and $\vert u_2\rangle$, respectively. 
We perform a $\pi$-rotation on spins on only even (or odd) sites of bipartite square lattice. It finally leads to an effective Hamiltonian for $J_2<J_1$ as
\begin{eqnarray}
\nonumber\mathcal{H}^{eff}&=& J_1^\prime\displaystyle\sum_{\langle I,J \rangle}{\tau_I^x\tau_J^x}
                            +  J_2^\prime\displaystyle\sum_{\langle\langle I,J \rangle\rangle}{\tau_I^x\tau_J^x}-\Gamma^\prime\displaystyle\sum_{I}{\tau_I^z},\\
\label{eqap7}                                             
\end{eqnarray}
where,
\begin{eqnarray}
\nonumber\frac{J_2^\prime}{J_1^\prime}&=&\frac{1}{2}\frac{J_2}{J_1},\\
\frac{\Gamma^{\prime}}{J_{1}^{\prime}}&=&\frac{\epsilon_2 - \epsilon_1}{2 \alpha^{2}}, \quad (J_2<J_1).
\label{eqap8}    
\end{eqnarray}
Similar procedure is also done for the case $J_2>J_1$.

%

\end{document}